\documentclass[%
 reprint,
superscriptaddress,
amsmath,amssymb,
aps,
]{revtex4-2}
\usepackage{amsmath} 
\usepackage{float}
\usepackage{graphicx}
\usepackage{dcolumn}
\usepackage{bm}
\usepackage{subcaption}
\usepackage{caption}
\usepackage{textcomp} 
\usepackage{xcolor}
\usepackage{gensymb}

\usepackage{hyperref}       

\hypersetup{
    colorlinks=true,        
    linkcolor=blue,         
    citecolor=blue,          
    urlcolor=magenta        
}

\begin{document}

\preprint{APS/123-QED}
\title{Auxiliary-Channel-Assisted Cross-Talk Noise Removal in LISA Pathfinder}

\author{Hui Sun}
\email{sunhui22@mails.ucas.ac.cn}
\affiliation{University of Chinese Academy of Sciences (UCAS), Beijing 100049, China
}%
\affiliation{International Centre for Theoretical Physics Asia-Pacific, UCAS, Beijing 100190, China
}%
\affiliation{Taiji Laboratory for Gravitational Wave Universe (Beijing/Hangzhou), UCAS, Beijing 100190, China
}

\author{Geyu Qin}
\affiliation{University of Chinese Academy of Sciences (UCAS), Beijing 100049, China
}%
\affiliation{International Centre for Theoretical Physics Asia-Pacific, UCAS, Beijing 100190, China
}%
\affiliation{Taiji Laboratory for Gravitational Wave Universe (Beijing/Hangzhou), UCAS, Beijing 100190, China
}

\author{Jibo He} 
\email{jibo.he@ucas.ac.cn}
\affiliation{University of Chinese Academy of Sciences (UCAS), Beijing 100049, China
}%
\affiliation{International Centre for Theoretical Physics Asia-Pacific, UCAS, Beijing 100190, China
}%
\affiliation{Taiji Laboratory for Gravitational Wave Universe (Beijing/Hangzhou), UCAS, Beijing 100190, China
}
\affiliation{Hangzhou Institute for Advanced Study, UCAS, Hangzhou 310024, China
}

\date{\today}

\begin{abstract}
LISA Pathfinder (LPF) is the technology demonstration mission for the future Laser Interferometer Space Antenna (LISA). Besides the science interferometric channel, LPF is equipped with numerous auxiliary channels that monitor the instrument and environment disturbances, some of which contain information correlated with noise in the science channel. In this work, we investigate cross-talk noise subtraction in LPF using a frequency-domain transfer-function approach, in which the coherence between the science channel and auxiliary channels is used to estimate and remove correlated noise. The effectiveness of this method is first validated using simulated science and auxiliary channel data with a common disturbance. The simulation shows that the subtraction performance strongly depends on the auxiliary-channel noise level, providing a practical framework for determining the auxiliary-channel noise requirement needed to achieve a desired subtraction performance. This method is then used to analyze 9 auxiliary channels in publicly available LPF telemetry data. It is found that for a 13-day dataset taken under stable conditions, the transfer functions estimated from the whole interval or separately for each day yield nearly identical subtraction performance over the frequency range under consideration. In the cross-talk dominated frequency band from $1\times10^{-2}$ to $6\times10^{-2}\,\mathrm{Hz}$, this method achieves cross-talk noise suppression comparable to the standard fitting based subtraction. Above $6\times10^{-2}\,\mathrm{Hz}$, it further suppresses the residual noise by avoiding the introduction of auxiliary-channel readout noise associated with the fitting procedure, resulting in better performance than the standard pipeline. We also assess the effect of glitch subtraction on the cross-talk subtraction results.

\end{abstract}

\maketitle


\section{introduction}

Since the first direct detection of gravitational waves (GW) by the Laser Interferometer Gravitational-Wave Observatory (LIGO) \cite{abbott2016observation}, GW has emerged as a powerful tool for studying compact objects and the universe, with a growing catalog of detected events~\cite{abbott2019gwtc, abbott2021gwtc, abbott2023gwtc, abac2026gwtc}. Extending GW observations to the millihertz band requires space-based detectors with sufficiently long interferometric baselines, for which several mission concepts have been proposed, including the Laser Interferometer Space Antenna (LISA) \cite{amaro2017laser}, Taiji~\cite{hu2017taiji}, and TianQin~\cite{luo2016tianqin}. To validate the key technologies required for these missions, dedicated precursor missions have been conducted, including LISA Pathfinder (LPF)~\cite{PhysRevLett.116.231101, PhysRevLett.120.061101}, Taiji-1~\cite{taiji2021china, cai2021satellite}, and TianQin-1~\cite{luo2020first, xiao2022drag}. In this study, we focus on LPF, for which telemetry data are publicly available. 

The science observable in LPF is the residual differential acceleration measured between the two test masses along the interferometric $x$-axis. In addition to the primary science interferometric channel, GW detectors are generally equipped with numerous auxiliary channels that monitor instrumental conditions and environmental disturbances \cite{acernese2023virgo, allocca2020interferometer}. Measurements from the science channel can be affected by various noise sources, some of which may simultaneously couple into auxiliary channels. In LPF, these noise sources include cross-talk noise, actuation noise, magnetic noise, shot noise, and Brownian noise \cite{castelli2020lisa, armano2021sensor}. While not all of these noise contributions are observable in auxiliary channels, some can simultaneously affect both the science and auxiliary channels, producing measurable correlations. Studying such correlations can therefore help identify common disturbance contributions and facilitate their subtraction from the science measurements. 

Among the auxiliary channels available in LPF, this work focuses primarily on those provided by the Optical Metrology Subsystem (OMS) \cite{armano2021sensor} and the Gravity Reference Sensor (GRS) \cite{armano2018calibrating}, which probe the dynamical state of the instrument. The OMS, a high-precision laser interferometric readout system, continuously measures the positions and orientations of the two test masses (TMs) relative to the spacecraft along six degrees of freedom ($x$, $\eta$, and $\phi$ for each TM). The GRS employs capacitive sensing to monitor the positions and orientations of the two TMs relative to the spacecraft along twelve degrees of freedom ($x$, $y$, $z$, $\theta$, $\eta$, and $\phi$ for each TM). In addition, LPF includes the Data and Diagnostics Subsystem (DDS) \cite{canizares2009diagnostics, armano2025magnetic}, which provides environmental monitoring through a temperature measurement subsystem \cite{armano2019temperature}, a magnetic diagnostics subsystem \cite{armano2020spacecraft}, and a radiation monitor \cite{armano2018characteristics}. 

These auxiliary channels have already demonstrated their value in LPF data analysis.~For example, coincident glitches observed in the science channel and OMS and GRS auxiliary channels have been used to investigate the physical origins of transient disturbances~\cite{armano2022transient}. Correlations between the LPF science channel and microthruster, thermometer, and magnetometer readouts have also been exploited to estimate the instrumental and environmental noise budget expected for LISA \cite{PhysRevD.106.063025}. These studies indicate that auxiliary channels can provide important information about the origin and coupling mechanisms of noise in the science measurements. Beyond noise identification, correlations between science and auxiliary channels can also be exploited for noise subtraction. However, the usefulness of an auxiliary channel in noise studies does not necessarily imply that it can enable effective correlated-noise subtraction.

In ground-based GW detectors, a variety of methods have been developed and successfully applied to subtract correlated noise using auxiliary-channel information. For example, a Wiener Filter \cite{wiener1949extrapolation} has been used in LIGO to mitigate laser noise arising from optical component vibrations induced by water-cooling systems~\cite{driggers2019improving}. In KAGRA, an Independent Component Analysis (ICA) has been employed to suppress suspension control noise and acoustic noise contributions \cite{abe2023noise}. Deep learning methods have also been explored for removing nonlinear noise couplings in LIGO~\cite{ormiston2020noise, saleem2024demonstration} and Virgo~\cite{kiendrebeogo2025application}.

In this work, we investigate correlations between the LPF science channel and a set of auxiliary channels from the OMS and GRS. We employ a frequency-domain transfer-function method for correlated-noise subtraction~\cite{allen1999automaticcrosstalkremovalmultichannel, davis2019improving}. Using controlled simulations with different auxiliary-channel noise levels, we first establish a practical framework that relates auxiliary-channel noise level to achievable subtraction performance, thereby providing quantitative guidance for auxiliary-channel noise requirement design in future space-based GW detectors. We then apply the same method to LPF telemetry data and demonstrate significant cross-talk noise subtraction. 

Our choice of the OMS and GRS channels is motivated by the established contribution of cross talk to the LPF noise budget. Previous studies have identified the coupling of spacecraft motion into the longitudinal readout as the origin of a noise bulge between 20 and 200 mHz, which limited the mission performance before subtraction~\cite{Wanner_2017, armano2023tilt}. We therefore focus on the OMS and GRS position and attitude readouts associated with the motion entering the standard cross-talk subtraction model~\cite{Wanner_2017}. 
For comparison, the magnetic contribution during the February 2017 noise run was estimated to account for approximately $1.5\%$ of the total acceleration-noise power at $0.1\,\mathrm{mHz}$~\cite{armano2025magnetic}. In our exploratory subtraction tests, the thermometer and magnetometer readouts did not yield a clear reduction in the science-channel noise; we therefore do not include them in the main subtraction analysis.

The remainder of this paper is organized as follows. In Sec.~\ref{sec:meth}, we present the theoretical framework of transfer-function-based noise subtraction, including both the two-channel and multichannel formulations. In Sec.~\ref{sec:sim}, we validate the method using simulated data and investigate the dependence of the subtraction performance on the auxiliary-channel noise level. In Sec.~\ref{sec:lpf}, we apply the proposed method to publicly available LPF telemetry data and compare its performance with the standard fitting-based cross-talk subtraction~\cite{Wanner_2017}. Finally, conclusions are presented in Sec.~\ref{sec:conclu}.

\section{Transfer-function-based noise subtraction}
\label{sec:meth}

This work exploits the coherence between the science channel and auxiliary channels to estimate and subtract correlated noise from the science channel. A frequency-dependent transfer function is constructed to estimate the correlated noise contribution from the auxiliary channels. The optimal transfer function is obtained by minimizing the residual power spectral density (PSD) of the science channel after subtraction. This section first introduces the definition of coherence, derives the optimal transfer function for the two-channel case, and then generalizes the method to multiple auxiliary channels.

\subsection{Coherence}

To quantify the linear correlation between the science channel and an auxiliary channel, we use the frequency-domain coherence, defined as 
\begin{equation}
\begin{aligned}
\rho_{xy}(f)
&=\frac{\left|\left\langle
\mathbf{X}^*(f)\mathbf{Y}(f)
\right\rangle\right|^2}
{\left\langle|\mathbf{X}(f)|^2\right\rangle
 \left\langle|\mathbf{Y}(f)|^2\right\rangle} \\
&=\frac{|P_{xy}(f)|^2}{P_{xx}(f)P_{yy}(f)}. 
\label{eq:coh}
\end{aligned}
\end{equation}
Here, $^*$ denotes complex conjugation, $|\cdot|$ denotes the complex modulus, and $\langle\cdot\rangle$ denotes averaging over data segments. Throughout this paper, the cross power spectral density (CSD) is defined as
\begin{equation}
P_{ab}(f)=\left\langle \mathbf{A}^*(f)\mathbf{B}(f)\right\rangle,
\label{eq:csd}
\end{equation}
where $\mathbf{A}(f)$ and $\mathbf{B}(f)$ denote the Fourier transforms of the corresponding data segments of the time-domain signals $a(t)$ and $b(t)$, respectively. The PSD is therefore a special case of the CSD, $P_{aa}(f)=\left\langle|\mathbf{A}(f)|^2\right\rangle$. In this work, all CSDs are estimated using Welch's method~\cite{w1161901} with a Blackman-Harris spectral window. Unless otherwise stated, the data are analyzed using segments of duration $T_{\rm seg}=T/100$, where $T$ denotes the total duration of the analyzed data.
Adjacent segments are overlapped by $50\%$.

We use the amplitude spectral density (ASD), defined as the square root of the PSD, to display signals and their noise components. Coherence and transfer functions are calculated from PSDs and CSDs, whereas the fractional contribution of a noise component to the total spectral power is expressed as a PSD ratio. The PSD has units of the squared signal unit per hertz, whereas the ASD has units of the signal unit per square root of hertz.

The coherence satisfies $0\le\rho_{xy}(f)\le1$. Unity coherence indicates that the spectral components of the two signals are perfectly linearly related, while zero coherence indicates that they are linearly uncorrelated. A higher coherence indicates that the auxiliary channel contains more information about the correlated noise in the science channel, enabling more effective estimation and subtraction of the correlated noise. 

\subsection{Two-channel noise subtraction}

Consider a science channel $\mathbf{X}(f)$ and an auxiliary channel $\mathbf{Y}(f)$. The residual signal after subtraction is
\begin{equation}
\mathbf{X'}(f)=\mathbf{X}(f)-T(f)\mathbf{Y}(f), 
\end{equation}
where $T(f)$ is the transfer function to be determined. The corresponding residual PSD is
\begin{equation}
P_{x'x'}(f)=\left\langle
\left[\mathbf{X}(f)-T(f)\mathbf{Y}(f)\right]^*
\left[\mathbf{X}(f)-T(f)\mathbf{Y}(f)\right]
\right\rangle.
\end{equation}
The optimal transfer function is obtained by minimizing $P_{x'x'}(f)$. The variation of the residual PSD is
\begin{equation}
\begin{aligned}
&\delta P_{x'x'}(f)\\
&= \left\langle-\left[\delta T(f)\mathbf{Y}(f)\right]^*
\left[\mathbf{X}(f)-T(f)\mathbf{Y}(f)\right]\right\rangle\\
&\quad +\left\langle-\left[\mathbf{X}(f)-T(f)\mathbf{Y}(f)\right]^*
\left[\delta T(f)\mathbf{Y}(f)\right]\right\rangle\\
&= -2\,\mathrm{Re}\!\left\{\delta T(f)\cdot
\left\langle
\left[\mathbf{X}(f)-T(f)\mathbf{Y}(f)\right]^*
\mathbf{Y}(f)
\right\rangle
\right\}. 
\end{aligned}
\end{equation}
Setting the variation to zero gives
\begin{equation}
\left\langle\left[\mathbf{X}(f)-T(f)\mathbf{Y}(f)\right]^*\mathbf{Y}(f)\right\rangle =0, 
\end{equation}
from which the optimal transfer function is obtained as
\begin{equation}
T(f)=\frac{\left\langle\mathbf{Y}^*(f)\mathbf{X}(f)\right\rangle}
{\left\langle\mathbf{Y}^*(f)\mathbf{Y}(f)\right\rangle}. 
\end{equation}
Substituting the optimal transfer function into the expression for $P_{x'x'}(f)$, the residual PSD can be written in terms of the original science-channel PSD as

\begin{equation}
\begin{aligned}
&P_{x'x'}(f)\\
&=\left\langle|\mathbf{X}(f)|^2\right\rangle
-T(f)\left\langle\mathbf{X}^*(f)\mathbf{Y}(f)\right\rangle\\
&\quad
-T(f)^*\left\langle\mathbf{Y}^*(f)\mathbf{X}(f)\right\rangle
+|T(f)|^2\left\langle|\mathbf{Y}(f)|^2\right\rangle\\
&=\left\langle|\mathbf{X}(f)|^2\right\rangle
-\frac{\left|
\left\langle\mathbf{Y}^*(f)\mathbf{X}(f)\right\rangle
\right|^2}
{\left\langle|\mathbf{Y}(f)|^2\right\rangle}\\
&= P_{xx}(f)\left[1-\rho_{xy}(f)\right].
\end{aligned}
\end{equation}

This result shows that the residual PSD depends on the coherence. A higher coherence leads to greater subtraction efficiency. When $\rho_{xy}=1$, the residual noise can be completely removed, yielding zero residual PSD. In contrast, when $\rho_{xy}=0$, the auxiliary channel contains no information about the science channel, and no noise subtraction is possible.

\subsection{Multi-channel noise subtraction}

In realistic scenarios, multiple auxiliary channels are often available, each of which may contain partial information about the noise in the science channel. This motivates extending the two-channel formulation to the multi-channel case.

Suppose that $N$ auxiliary channels, denoted by $\mathbf{Y}_j(f)$ $(j=1,\cdots,N)$, are available. Let $T_j(f)$ denote the transfer function associated with the $j$th auxiliary channel. The residual science channel is expressed as 
\begin{equation}
\mathbf{X'}(f)=\mathbf{X}(f)-\sum_{j=1}^N T_j(f)\mathbf{Y}_j(f).
\end{equation}
The variation of the residual PSD is
\begin{widetext}
\begin{equation}
\begin{aligned}
&\delta P_{x'x'}(f)\\
&=\delta
\left\langle
\left[\mathbf{X}(f)-\sum_{j=1}^N T_j(f)\mathbf{Y}_j(f)\right]^*
\left[\mathbf{X}(f)-\sum_{k=1}^N T_k(f)\mathbf{Y}_k(f)\right]
\right\rangle\\
&=
\left\langle
-\sum_{j=1}^N
\left[\delta T_j(f)\mathbf{Y}_j(f)\right]^*
\left[\mathbf{X}(f)-\sum_{k=1}^N T_k(f)\mathbf{Y}_k(f)\right]
\right\rangle+
\left\langle
-\sum_{k=1}^N
\left[\mathbf{X}(f)-\sum_{j=1}^N T_j(f)\mathbf{Y}_j(f)\right]^*
\left[\delta T_k(f)\mathbf{Y}_k(f)\right]
\right\rangle\\
&=
-2\,\mathrm{Re}\!\left\{
\sum_{k=1}^N
\delta T_k(f)
\left\langle
\left[\mathbf{X}(f)-\sum_{j=1}^N T_j(f)\mathbf{Y}_j(f)\right]^*
\mathbf{Y}_k(f)
\right\rangle
\right\}.
\end{aligned}
\end{equation}
\end{widetext}
Setting $\delta P_{x'x'}(f)$ to zero gives
\begin{equation}
\left\langle\left[\mathbf{X}(f)-\sum_{j=1}^NT_j(f)\mathbf{Y}_j(f)\right]^*\mathbf{Y}_k(f)\right\rangle=0,
\end{equation}
for every $k=1,\cdots,N$. The optimal transfer functions can be written in matrix form as
\begin{equation}
T_j(f)=\sum_{k=1}^N\left(R^{-1}\right)_{jk}(f)\cdot P_{k}(f),\quad\mathrm{for~}j=1,2,\cdots,N.
\label{eq:tf2}
\end{equation}
Here, 
\begin{equation}
R_{jk}(f)=
\left\langle
\mathbf{Y}_{j}(f)^*\mathbf{Y}_{k}(f)
\right\rangle
\end{equation}
is the correlation matrix among the auxiliary channels, and 
\begin{equation}
P_{k}(f)=
\left\langle
\mathbf{Y}_{k}(f)^*\mathbf{X}(f)
\right\rangle
\end{equation}
is the cross-correlation vector between the auxiliary channel and the science channel.

Equation~\eqref{eq:tf2} provides a general solution for transfer-function-based multi-channel noise subtraction. At each frequency, the optimal transfer functions can be obtained by computing the CSDs among the auxiliary channels and the science channel. The performance of this method will be demonstrated using both simulated data and publicly available LPF telemetry data in the following sections.

\begin{figure}
\centering
\begin{subfigure}{\columnwidth}
    \includegraphics[width=\linewidth]{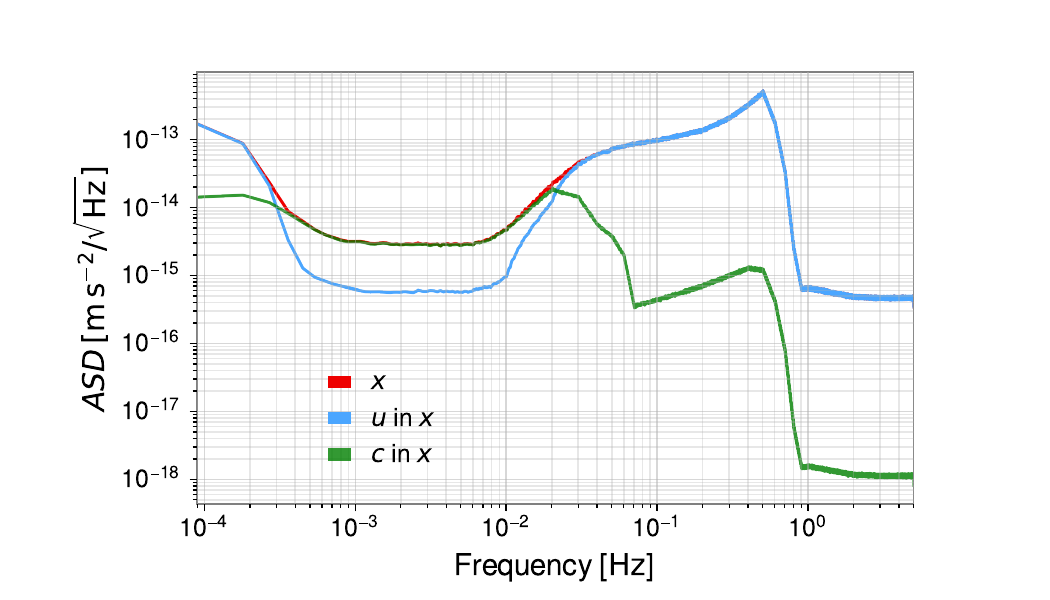}
    \caption{Simulated science channel x and its components.}
\end{subfigure}
\vspace{1mm}
\begin{subfigure}{\columnwidth}
    \includegraphics[width=\linewidth]{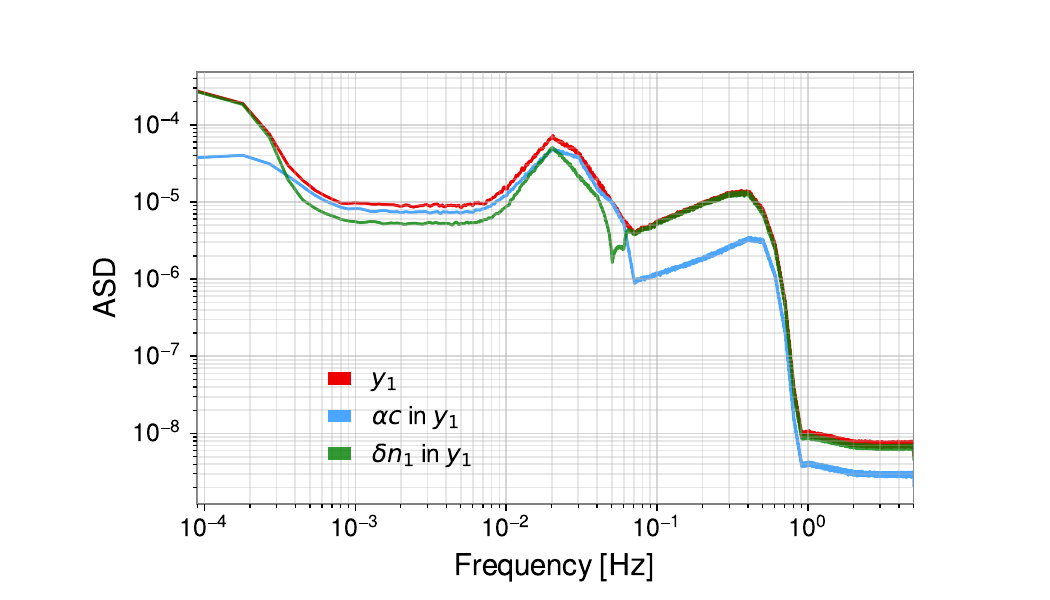}
    \caption{Simulated auxiliary channel $y_1$ and its components.}
\end{subfigure}
\caption{\label{fig:xy1comp} ASDs of the simulated science channel $x$ and a representative auxiliary channel $y_1$ at $\delta = 1.0$, together with their component decomposition. The science channel $x$ is composed of the science signal $u(t)$ and a common disturbance $c(t)$, while the auxiliary channel $y_1$ consists of a scaled common disturbance term $\alpha c(t)$ and a noise term $\delta n_1(t)$. The ASDs in panel (a) have units of $\mathrm{m\,s^{-2}\,Hz^{-1/2}}$. In panel (b), the ASDs are expressed in units of the auxiliary observable divided by the square root of hertz.}
\end{figure}

\begin{figure}
\includegraphics[width=8.5cm]{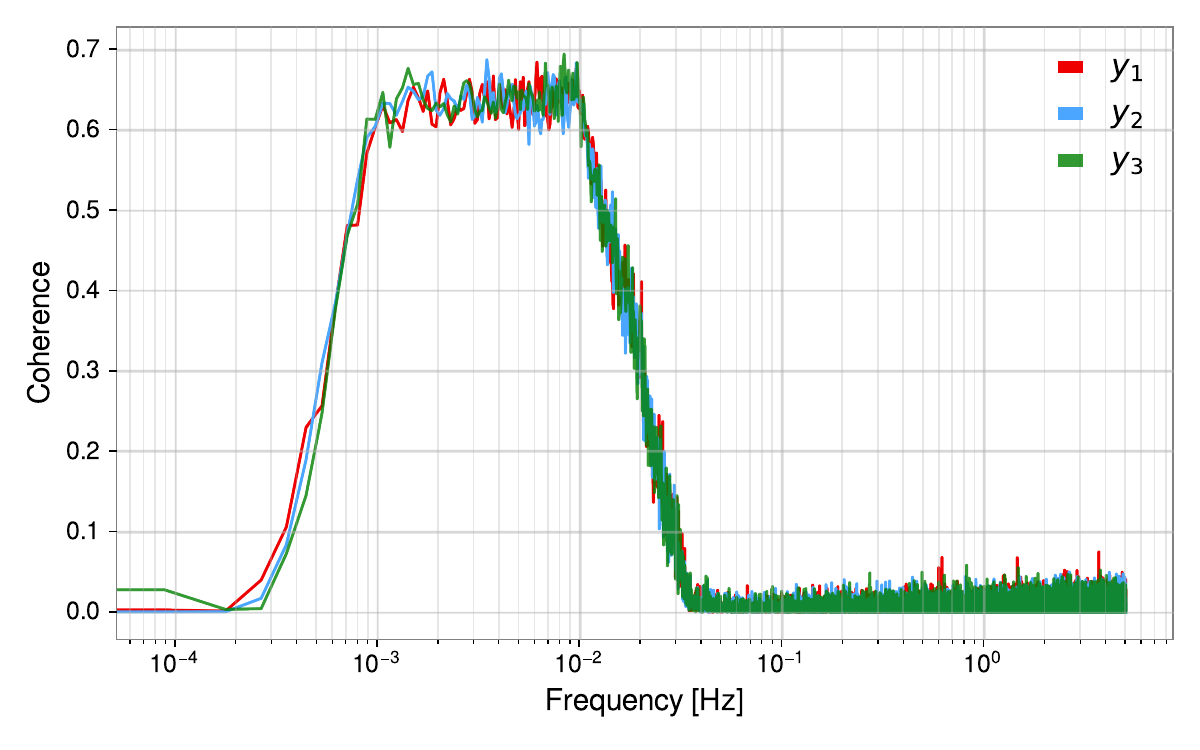}
\captionsetup{justification=raggedright,singlelinecheck=false}
\caption{\label{fig:coh0} Coherence between the simulated science channel $x$ and the auxiliary channels $y_i~(i=1,2,3)$ for $\delta=1.0$. A high level of coherence is observed in the approximately $1~\mathrm{mHz}$ to $10~\mathrm{mHz}$ band, indicating a strong coupling of the common disturbance across channels. }
\end{figure}

\begin{figure*}[t]
\centering


\begin{subfigure}{0.33\textwidth}

    \centering

    \includegraphics[width=\linewidth]{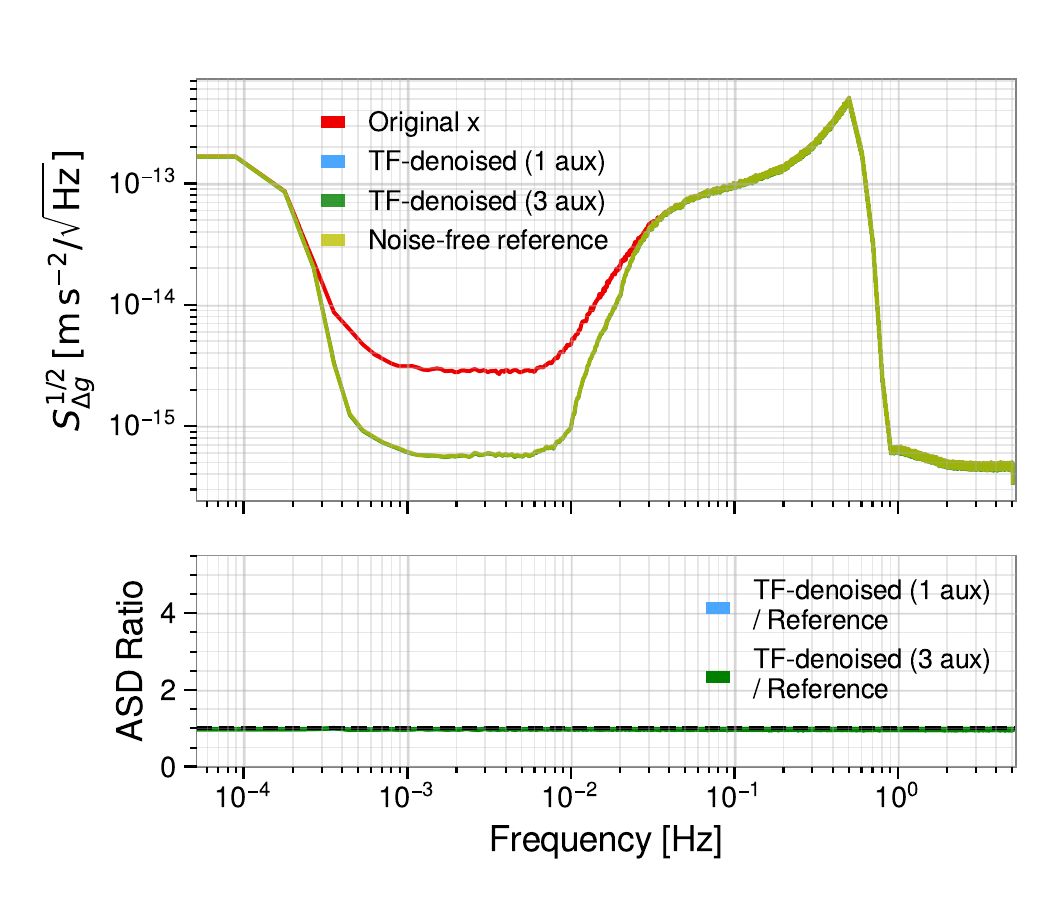}

    \caption{$\delta=0.01$}\label{fig:asdr-a}

\end{subfigure}\hfill
\begin{subfigure}{0.33\textwidth}

    \centering

    \includegraphics[width=\linewidth]{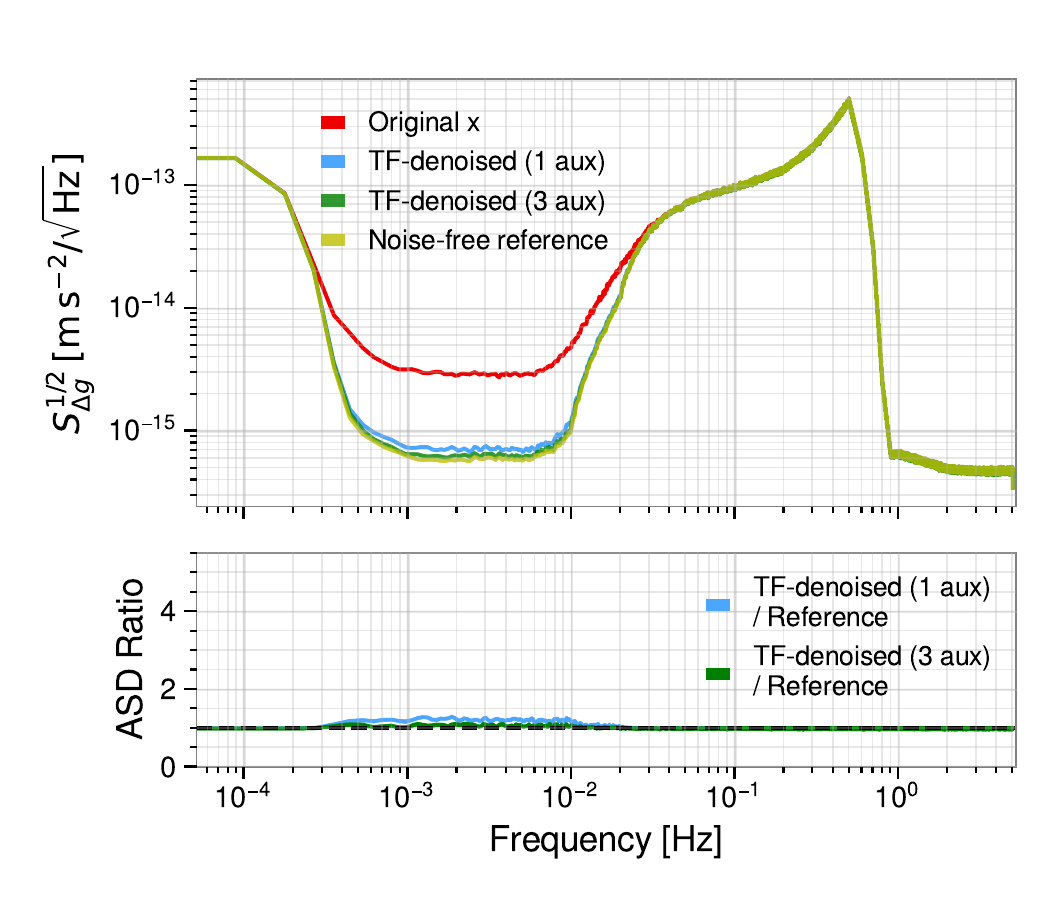}

    \caption{$\delta=0.2$}

\end{subfigure}\hfill
\begin{subfigure}{0.33\textwidth}

    \centering

    \includegraphics[width=\linewidth]{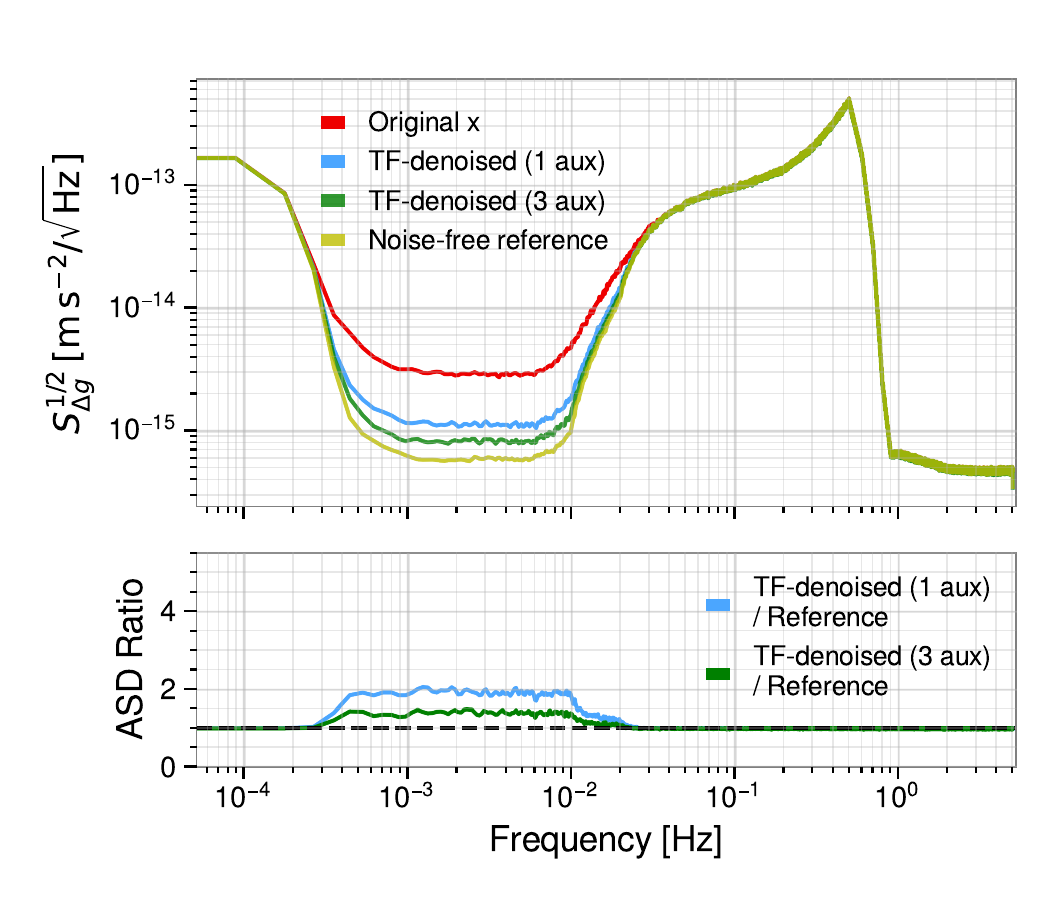}

    \caption{$\delta=0.5$}

\end{subfigure}

\vspace{2mm}


\begin{subfigure}{0.33\textwidth}

    \centering

    \includegraphics[width=\linewidth]{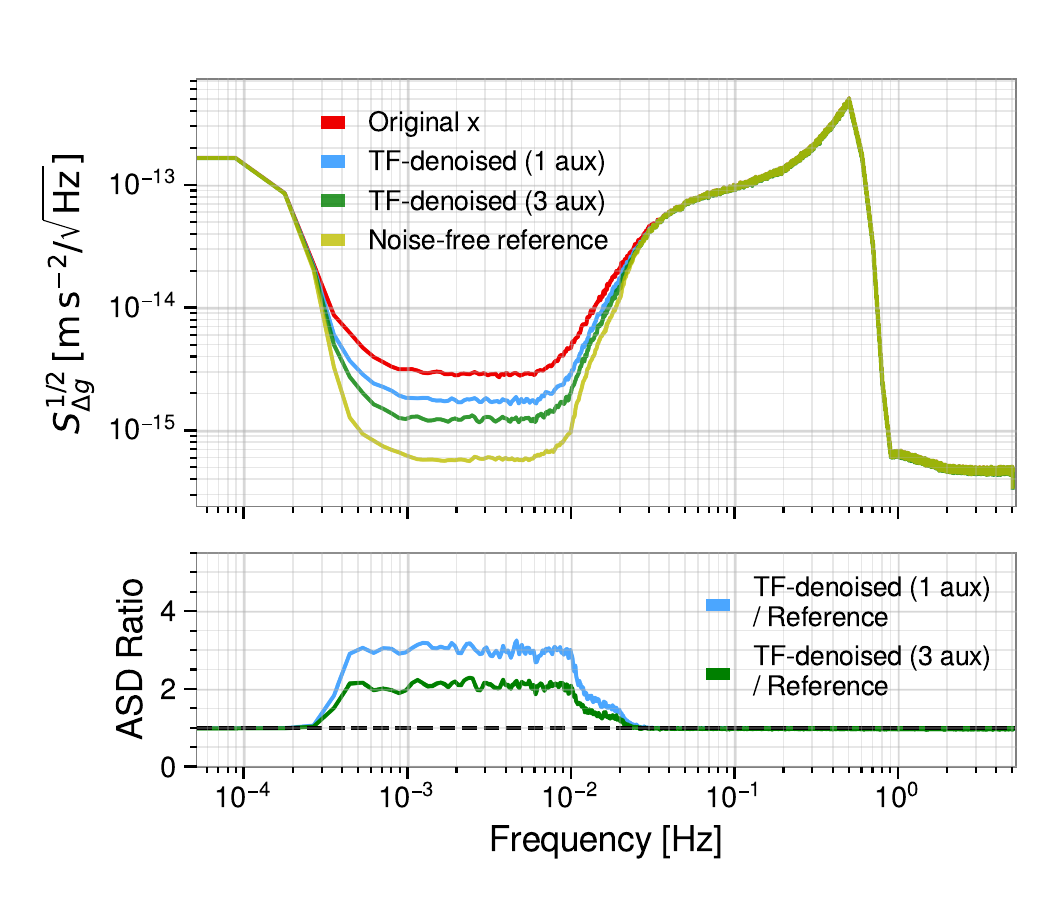}

    \caption{$\delta=1.0$}

\end{subfigure}\hfill
\begin{subfigure}{0.33\textwidth}

    \centering

    \includegraphics[width=\linewidth]{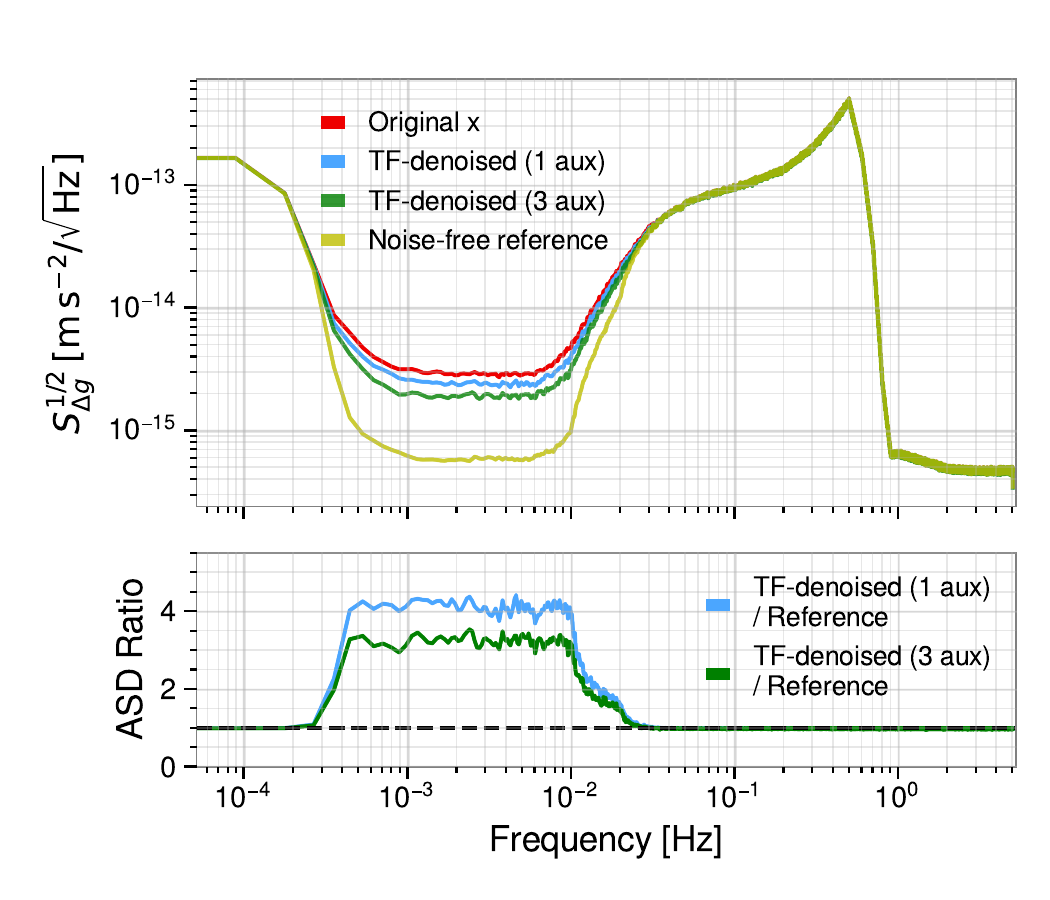}

    \caption{$\delta=2.0$}

\end{subfigure}\hfill
\begin{subfigure}{0.33\textwidth}

    \centering

    \includegraphics[width=\linewidth]{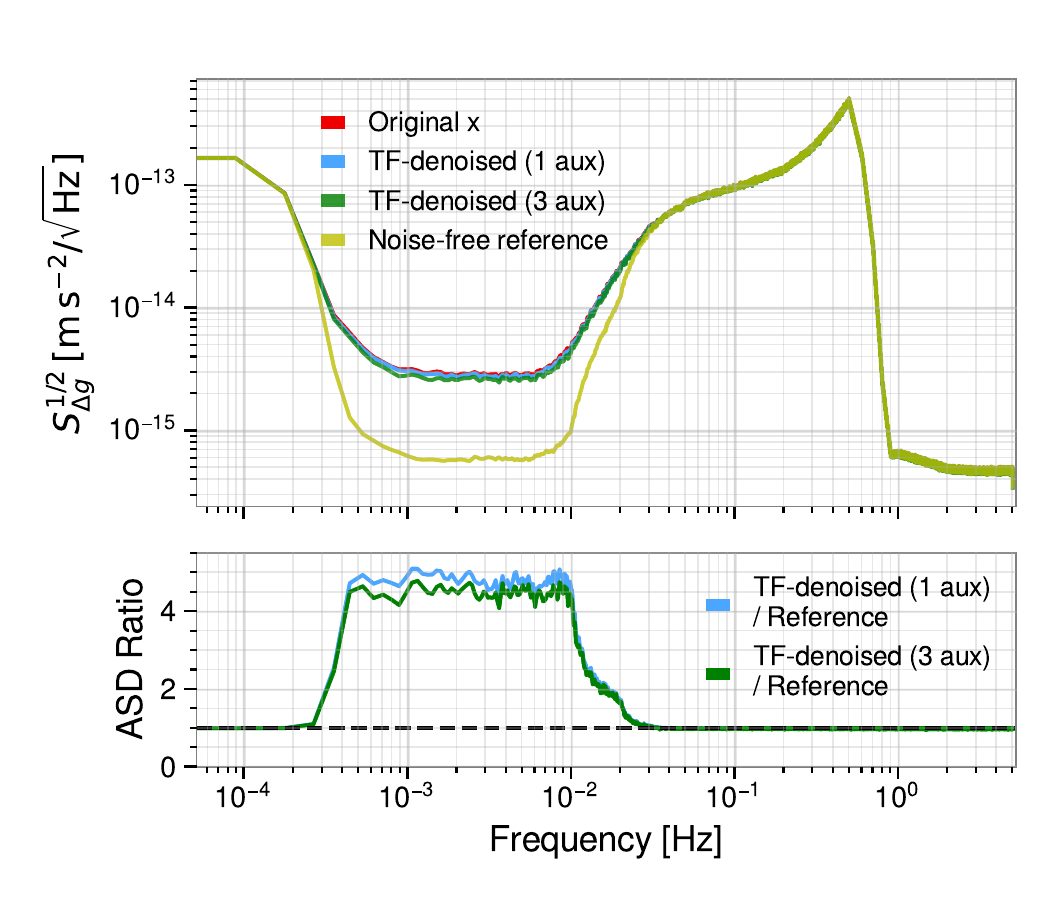}

    \caption{$\delta=5.0$}\label{fig:asdr-f}

\end{subfigure}

\caption{\label{fig:asdr} Transfer-function-based noise subtraction results for simulated data with different auxiliary channel noise levels $\delta$. Each panel corresponds to a different value of $\delta$. The upper subpanel shows the ASD of the original science channel before subtraction (red), the subtraction result using a single auxiliary channel $y_1$ (blue), the joint subtraction result using three auxiliary channels (green), and the noise-free reference $u(t)$ (yellow). The lower subpanel shows the ratios between each subtraction result and the noise-free reference ASD. As $\delta$ increases, the subtraction performance degrades. }
\end{figure*}

\section{Simulation and Noise Subtraction}
\label{sec:sim}

To validate the proposed transfer-function-based noise subtraction method, we construct a simulated multi-channel dataset consisting of one science channel and three auxiliary channels. Besides validating the method, the simulation is designed to investigate how the auxiliary-channel noise level affects the achievable subtraction performance. Establishing this relationship provides a practical basis for determining auxiliary-channel noise requirements during the design of future space-based GW detectors. 

The science channel is modeled as a linear combination of the science signal and a common disturbance component 
\begin{equation}
x(t)=u(t)+c(t),
\end{equation}
where $u(t)$ denotes the science signal to be preserved in the science channel, and $c(t)$ represents the common disturbance coupled into both the science channel and the auxiliary channels.

The $i$th auxiliary channel is constructed as 
\begin{equation}
y_i(t)=c'(t)+\delta n_i(t),
\end{equation}
where $c'(t)$ denotes a scaled version of the common disturbance, defined as $c'(t)=\alpha c(t)$, with a fixed scaling factor $\alpha=2.5\times10^{9}$. The factor $\alpha$ represents the effective coupling strength of the common disturbance between the auxiliary and science channels, reflecting the different response scales of the corresponding readout systems. $n_i(t)$ denotes an independent random noise process with the prescribed target amplitude spectral density (ASD), and $\delta$ is the amplitude coefficient of the auxiliary-channel noise $n_i(t)$. 

All simulated time series have a duration of 13 days and are sampled at $10~\mathrm{Hz}$, matching the LPF dataset analyzed in Sec.~\ref{sec:lpf}. Each process is generated from independent random variables uniformly distributed on the interval $[-1,1]$. The Fourier phases of the resulting time series are preserved, while the Fourier amplitudes are rescaled to the prescribed target ASD, thereby producing stationary random processes with the desired spectral shape. The science signal $u(t)$, the common disturbance $c(t)$, and the auxiliary-channel noise $n_i(t)$ are assigned three different target ASDs, with the auxiliary-channel noises $n_i(t)$ sharing the same target ASD. Since all processes are generated from independent random sequences, they are statistically independent of one another.

The simulated science channel $x$ represents residual acceleration. Accordingly, the ASDs of $x$, $u$, and $c$ have units of $\mathrm{m\,s^{-2}\,Hz^{-1/2}}$. The ASDs of $y_i$ and $\delta n_i$ have units determined by the auxiliary observable, for example, $\mathrm{m\,Hz^{-1/2}}$ for displacement or $\mathrm{T\cdot Hz^{-1/2}}$ for magnetic field. The scaling factor $\alpha$ accounts for the corresponding difference in units and response scale.

The dimensionless parameter $\delta$ controls the noise level of the auxiliary channels and therefore determines the degree of correlation between the science channel and the auxiliary channels. To investigate the influence of the auxiliary-channel noise on the subtraction performance, the following values are considered: 
\begin{equation}
\delta=0.01,\;0.2,\;0.5,\;1.0,\;2.0,\;5.0.
\end{equation}
As an example, Figure~\ref{fig:xy1comp} shows the ASDs of the science channel $x$ and a representative auxiliary channel $y_1$ for $\delta=1.0$, together with their individual components. The science channel consists of the science signal $u(t)$ and the common disturbance $c(t)$, while the auxiliary channel consists of the scaled common disturbance component $\alpha c(t)$ and the independent noise component $\delta n_1(t)$. 

Figure~\ref{fig:coh0} further presents the coherence between the science channel $x$ and the auxiliary channels $y_i~(i=1,2,3)$ for $\delta=1.0$. A high level of coherence is observed approximately over the frequency band from $1~\mathrm{mHz}$ to $10~\mathrm{mHz}$, indicating strong common-noise coupling across the channels and providing conditions for common disturbance estimation and subtraction using the proposed transfer-function method. 

The noise subtraction performance for different auxiliary-channel noise levels
is compared in Fig.~\ref{fig:asdr}\subref{fig:asdr-a}--\subref{fig:asdr-f}. The upper panels compare the ASDs of the original science channel, the subtraction results obtained using one and three auxiliary channels, and the noise-free reference signal $u(t)$, while the lower panels show the corresponding ASD ratios relative to the reference. When $\delta$ is small, the auxiliary channels provide an accurate estimate of the common disturbance, and the ASD of the denoised science channel closely approaches the noise-free reference $u(t)$. As $\delta$ increases, the auxiliary-channel noise progressively reduces the correlation with the science channel, leading to a gradual degradation of the subtraction performance. Meanwhile, the advantage of using three auxiliary channels over a single auxiliary channel becomes increasingly significant as $\delta$ increases.

We quantify the auxiliary channel's intrinsic noise $\delta n_i$ contribution as the ratio of its noise PSD to its total PSD. In the $1\text{--}5~\mathrm{mHz}$ band, Figure~\ref{fig:xy1comp} shows that $\delta=1.0$ corresponds to an auxiliary-channel noise $\delta n_1$ contribution of approximately 40\% of the total auxiliary-channel PSD. The results shown in Fig.~\ref{fig:asdr} indicate that, in the present simulation, the subtraction performance degrades significantly for $\delta \ge 1.0$. More generally, the present simulation framework provides a practical means of determining the auxiliary-channel noise requirement for a desired subtraction performance, provided that the contribution of the common disturbance to the science channel is known.

\section{Cross-talk noise subtraction in LPF data}
\label{sec:lpf}

To validate the effectiveness of the proposed transfer-function-based noise subtraction method in real space-based GW data, we apply it to publicly available telemetry from the LISA Pathfinder (LPF) mission. The data are taken from the LISA Pathfinder Legacy Archive~\cite{LPF_Legacy_Archive}. The LPF science channel records the residual differential acceleration between the two TMs along the sensitive $x$-axis. The science channel data are provided in processing levels from L0 to L4, with successive corrections applied for centrifugal force, cross-talk forces, Euler forces, and glitch noise, respectively \cite{armano2018calibrating}. All science-channel data products (L0–L4) are sampled at $10~\mathrm{Hz}$. 

In particular, Level 1 (L1) data correspond to residual differential acceleration measurements without cross-talk subtraction, while Level 2 (L2) data are the publicly released products obtained after applying the standard linear fitting-based cross-talk removal procedure. Therefore, the L1–L2 pair provides a natural dataset for evaluating cross-talk mitigation methods.

We analyze 13 consecutive days from 14 February 2017, 02:00 UTC, to 27 February 2017, 02:00 UTC. All L1, L2 data and auxiliary channels used here cover this same interval, during which the spacecraft and TMs remained under stable control without planned calibration excitations. Neither L1 nor L2 includes glitch subtraction, which is applied in the processing from L3 to L4. We first compare the two cross-talk subtraction methods without additional glitch subtraction and then assess how prior glitch subtraction affects the performance of the transfer-function-based method.

\subsection{Standard fitting-based cross-talk subtraction}

Cross-talk noise refers to spurious contributions in the residual differential acceleration readout of the science channel arising from couplings with non-scientific degrees of freedom. In the standard LPF data processing pipeline, cross-talk is estimated and subtracted using a linear parametric fitting procedure \cite{Wanner_2017}. In this approach, a cross-talk model is constructed from the TMs positions and orientations measurement auxiliary channels and removed from the L1 residual differential acceleration, yielding the L2 data product, 
\begin{equation}
    \begin{aligned}\Delta g_\mathrm{L2}(t)&=\Delta g_\mathrm{L1}(t)-\delta g_{\mathrm{xtalk}}(t).\end{aligned}
    \label{equ:l2}
\end{equation}
Here, $\Delta g_\mathrm{L1}(t)$ and $\Delta g_\mathrm{L2}(t)$ denote the residual acceleration in the L1 and L2 data products, respectively, and $\delta g_{\mathrm{xtalk}}(t)$ represents the estimated cross-talk contribution obtained from the fitting model. The standard fitting model is given by \cite{Wanner_2017}
\begin{equation}
    \begin{aligned}\delta g_\mathrm{xtalk}(t)=&b_1\ddot{\overline{\phi}}(t)+b_2\ddot{\overline{\eta}}(t)+b_3\ddot{\overline{y}}(t)\\&+b_4\ddot{\overline{z}}(t)+b_5\overline{y}(t)+b_6\overline{z}(t)+b_7\ddot{x_1}(t).\end{aligned}
\end{equation}
Here, $\phi$ and $\eta$ denote the TM attitude angles corresponding to rotations about the $z$-axis and $y$-axis (yaw and pitch), respectively. Subscripts 1 and 2 refer to the corresponding quantities of TM1 and TM2. The averaged quantities are defined as $\bar{\phi} = (\phi_1 + \phi_2)/2$, with analogous definitions for other barred variables. The operator $\ddot{(\cdot)}$ denotes the second derivative with respect to time. The auxiliary measurement channels entering the model include $\phi_1$, $\phi_2$, $\eta_1$, $\eta_2$, $y_1$, $y_2$, $z_1$, $z_2$, and $x_1$.

The OMS provides high-precision measurements of the TM position along the $x$ direction as well as the attitude angles $\eta$ and $\phi$. Therefore, the $x$, $\eta$, and $\phi$ degrees of freedom are taken from the OMS data. Since OMS does not provide measurements of the $y$ and $z$ directions, these degrees of freedom are instead obtained from the GRS capacitive readout system.

The first six terms in the model describe cross-talk contributions arising from linear couplings of motions along the orthogonal coordinates into the science channel, while the last term represents a leakage contribution of the spacecraft common-mode motion along the sensitive $x$-axis arising from a slight mismatch in the readout response between the two test masses, denoted by $\ddot{x}_1$. 

The ASDs of L1 and L2 over this time interval are shown as the blue and green curves, respectively, in Fig.~\ref{fig:denoising}. It is observed that the standard fitting method effectively suppresses the cross-talk bump in the frequency band $[1\times10^{-2}, 2\times10^{-1}]\,\mathrm{Hz}$. However, an increase of the ASD is observed at higher frequencies in the L2. This behavior is attributed to the introduction of readout noise from the auxiliary-channel capacitive sensors during the fitting-based subtraction procedure.

\subsection{Transfer-function-based cross-talk subtraction}

\begin{figure}
\includegraphics[width=8.5cm]{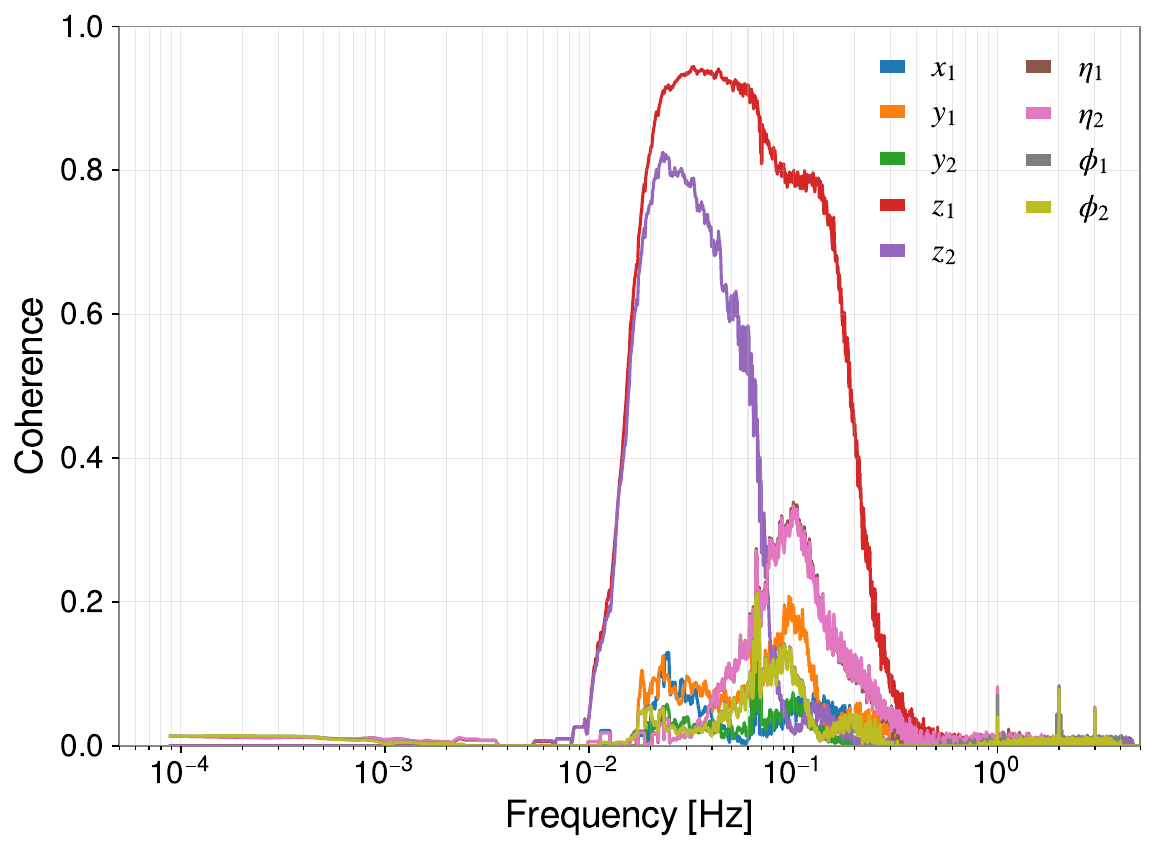}
\captionsetup{justification=raggedright,singlelinecheck=false}

\caption{\label{fig:cohz1} Coherence between the LPF science channel and 9 auxiliary channels in the data taken from 14 February 2017, 02:00 UTC, to 27 February 2017, 02:00 UTC. To improve visualization, the coherence spectra are smoothed using a 15-point moving average. The coherence exhibits significant variations among different auxiliary channels. Some channels show relatively strong coherence with the science channel, and even channels associated with the same degree of freedom can display distinct coherence characteristics, providing the basis for the multichannel noise subtraction presented in this work.}
\end{figure}

Figure~\ref{fig:cohz1} shows the coherence spectra between the LPF science channel (L1) and 9 auxiliary measurement channels. The auxiliary channels are sampled at the same rate as the science channel ($10~\mathrm{Hz}$), allowing the coherence and transfer functions to be computed directly without resampling. To improve readability, a 15-point moving average is applied to smooth the coherence spectra. It is observed that the coherence between the science channel and different auxiliary channels exhibits variations across both frequency and channel index. In particular, the $z_1$ and $z_2$ channels show relatively high coherence over a broad frequency band, while the remaining auxiliary channels exhibit weaker coherence overall.

\begin{figure}
\includegraphics[width=8.5cm]{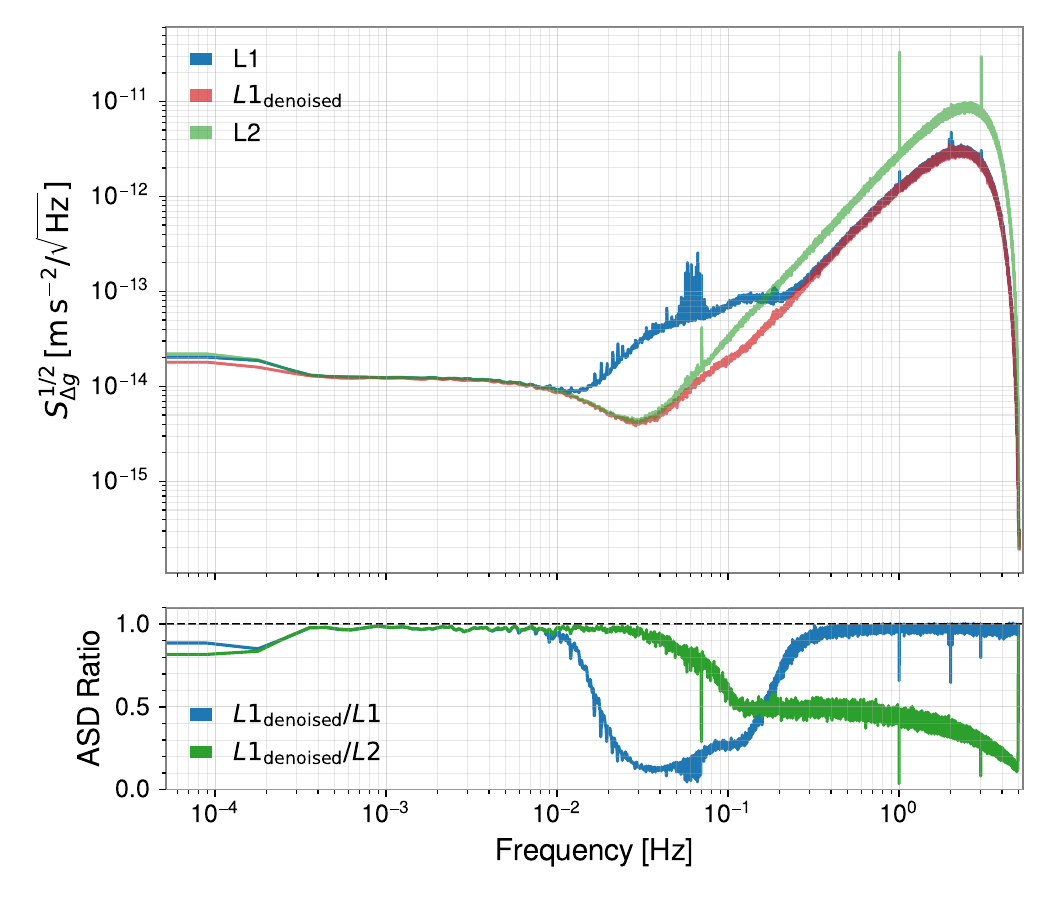}
\captionsetup{justification=raggedright,singlelinecheck=false}
\caption{\label{fig:denoising} ASDs of the LPF L1 science channel and the corresponding residual noise spectra obtained using two different noise subtraction methods, together with the associated ASD ratios, in the data taken from 14 February 2017, 02:00 UTC, to 27 February 2017, 02:00 UTC. The upper panel shows the ASD of the original L1 data (blue), the residual spectrum obtained using the transfer-function-based method (red), and the standard fitting-based L2 result (green). Compared with the fitting-based method, the transfer-function-based subtraction achieves further suppression of the residual noise spectrum in the high frequency band ($>6\times10^{-2}\,\mathrm{Hz}$). The lower panel shows the corresponding ASD ratios. The blue curve gives the ratio between the transfer-function-based result and the L1 ASD, and the green curve gives the ratio between the transfer-function-based result and the L2 ASD.}
\end{figure}

When the coupling is approximately linear and the instrument operates under stable conditions, the physical coupling response can be treated as approximately time invariant over the period analyzed. A single set of frequency-dependent transfer functions may therefore suffice to describe the correlated-noise contribution over this interval, without requiring daily updates. Previous LPF studies report approximately stationary acceleration noise over week-long measurement runs~\cite{armano2022transient}. 

\begin{figure}
\includegraphics[width=\columnwidth]{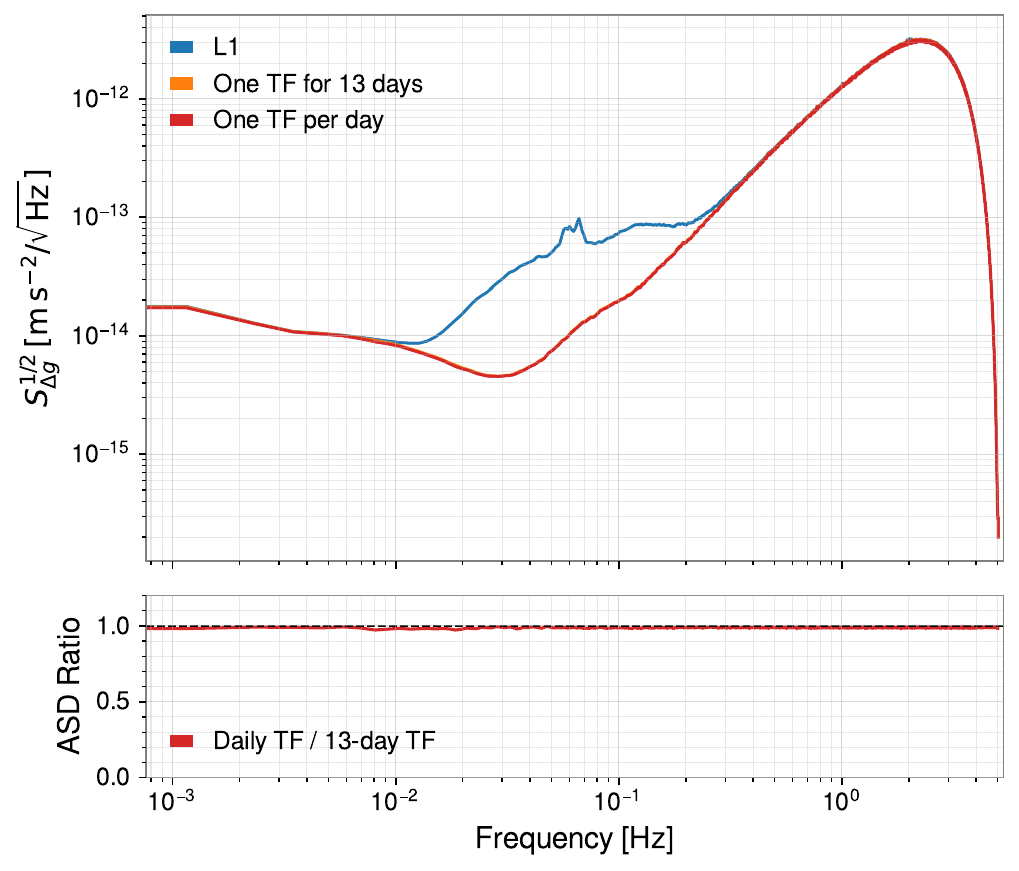}
\captionsetup{justification=raggedright,singlelinecheck=false}
\caption{Comparison of cross-talk subtraction using transfer functions estimated from the full 13-day dataset and separately for each day for the data taken from 14 February 2017, 02:00 UTC, to 27 February 2017, 02:00 UTC. The transfer functions are estimated using Blackman--Harris windows of duration one hundredth of a day with $50\%$ overlap, so the full-dataset estimates average over more segments than the daily estimates. The upper panel shows the original L1 ASD (blue) and the residual ASDs obtained using full-dataset estimates (orange) and daily estimates (red). The lower panel shows the ratio of the daily residual ASD to the full-dataset residual ASD.}
\label{fig:dailyTF}
\end{figure}

To assess whether daily updates are needed for this dataset, we compare one transfer function per auxiliary channel estimated from the full 13-day dataset with transfer functions estimated separately for each day, using the original L1 data. For this comparison, the spectral averages entering the transfer-function solution in Eq.~\eqref{eq:tf2} are computed using the same Blackman--Harris window duration, set to one hundredth of a day, with $50\%$ overlap. The full-dataset estimate therefore averages over more segments than each daily estimate, while both approaches have the same frequency resolution. As shown in Fig.~\ref{fig:dailyTF}, the two approaches yield nearly identical residual ASDs over the considered frequency range. The nearly identical subtraction performance supports the use of full-dataset transfer-function estimates. In the subsequent analysis, we estimate one transfer function per auxiliary channel from the full 13-day dataset, using $T_{\rm seg}=T/100$ with $T=13\,\mathrm{days}$ to extend the frequency coverage to lower frequencies.

The transfer-function-based noise subtraction method is applied to L1, and the resulting residual spectrum is shown as the red curve in Fig.~\ref{fig:denoising}.~It is observed that the transfer-function-based subtraction method achieves better performance than the fitting-based method in the frequency range above $6\times10^{-2}\,\mathrm{Hz}$. In addition, in the frequency band $[1\times10^{-2}, 6\times10^{-2}]\,\mathrm{Hz}$, it yields a comparable noise suppression performance to that of the standard fitting-based subtraction.

Previous studies have shown that TM realignment can suppress the high-frequency ASD increase observed in the L2 data after fitting-based cross-talk subtraction by reducing the contribution of position readout noise associated with TM misalignment~\cite{armano2023tilt, hartig2022tilt}. For the dataset considered here, the time interval is characterized by a relatively long elapsed period of approximately 8 months since the last TM alignment (25 June 2016)~\cite{hartig2022tilt}, making it representative of such a misalignment condition. However, TM realignment requires updating the drag-free attitude control system (DFACS) reference angles and must be performed online \cite{hartig2022tilt}. Therefore, for operational periods in which no recent alignment is available, the transfer-function-based subtraction method presented in this work provides an effective offline alternative.

\begin{figure}
\centering
\includegraphics[width=\columnwidth]{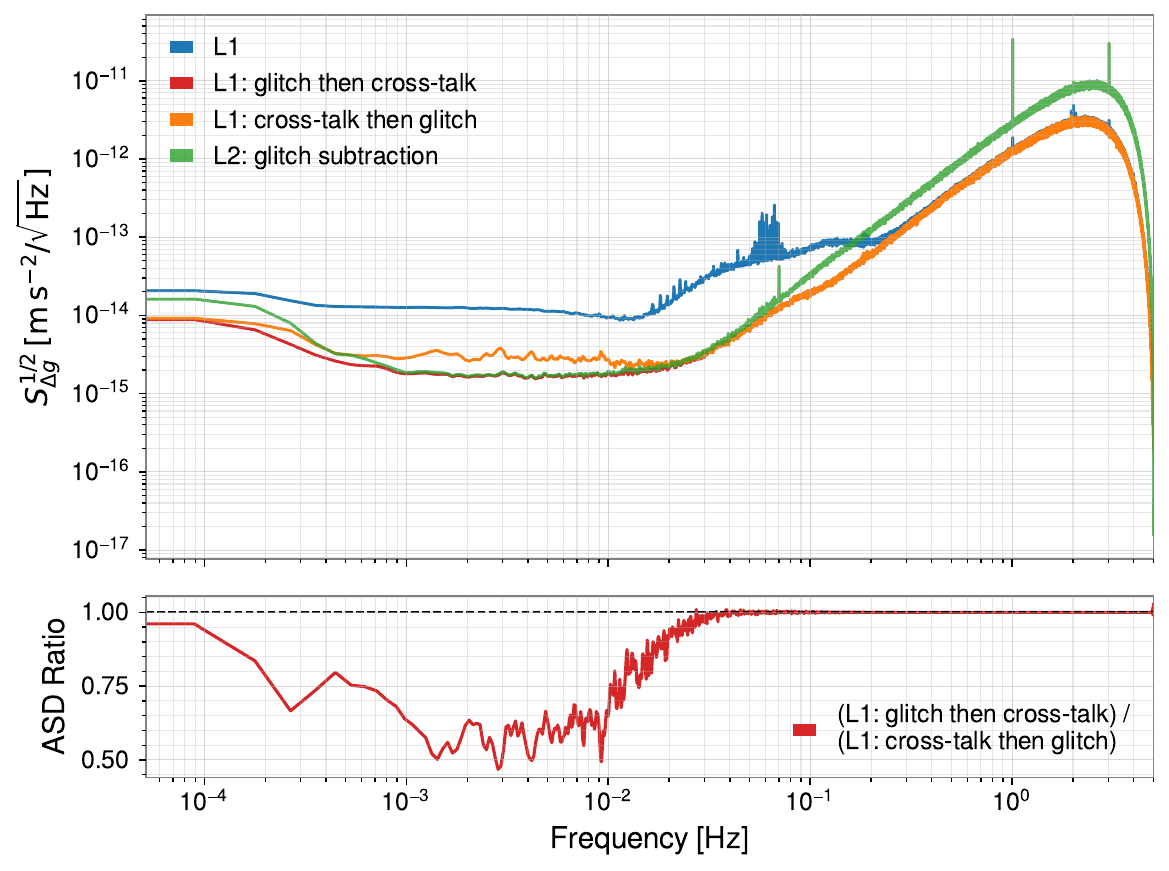}
\captionsetup{justification=raggedright,singlelinecheck=false}

\caption{\label{fig:glitchTF} Effect of the order of glitch subtraction and transfer-function-based cross-talk subtraction for the data taken from 14 February 2017, 02:00 UTC, to 27 February 2017, 02:00 UTC. The upper panel shows the original L1 ASD (blue), the residual ASD obtained by subtracting glitches before estimating the transfer functions and removing cross talk (red), the residual ASD obtained by estimating the transfer functions and removing cross talk before subtracting glitches (orange), and L2 after glitch subtraction (green). The lower panel shows the ratio of the red and orange ASDs.}
\end{figure}

\subsection{Glitch subtraction before transfer-function estimation}
\label{sec:glitch}

The comparisons above are performed without additional glitch subtraction. To assess whether glitches should be removed before applying the transfer-function-based cross-talk subtraction method, we compare results obtained with and without prior glitch subtraction. For a direct comparison of the final residuals, glitches are also fitted and subtracted after cross-talk subtraction in the latter case. The two procedures therefore differ in whether glitch subtraction is performed before or after transfer-function estimation and cross-talk subtraction. 

Glitches have been observed in LPF differential-acceleration data at a rate of about one per day~\cite{armano2022transient}. Using the analytic models described in Ref.~\cite{armano2022transient}, we fit glitches separately to the original L1 data and the L1 residual after transfer-function-based cross-talk subtraction, following the same fitting procedure in both cases. The fitted glitch contribution is subtracted from each corresponding input time series. In the glitch-first procedure, the transfer functions are estimated from L1 after glitch subtraction.

The red and orange curves in Fig.~\ref{fig:glitchTF} show the results obtained with glitch subtraction before and after transfer-function-based cross-talk subtraction, respectively; the lower panel shows their ASD ratio. Within the frequency range shown, this ratio is generally below unity for $f\lesssim4\times10^{-2}\,\mathrm{Hz}$. Thus, subtracting glitches after cross-talk subtraction does not fully recover the performance obtained when glitches are removed first. This difference may reflect the nonstationary nature of glitches: removing them first can improve the estimation and subtraction of correlated noise. 

For comparison with the standard fitting-based cross-talk subtraction method, we also fit and subtract glitches from L2, which has not undergone glitch subtraction because the glitch correction is applied in the processing from L3 to L4. The resulting ASD is shown in green in Fig.~\ref{fig:glitchTF}. The glitch-first transfer-function-based result retains its lower residual ASD above $6\times10^{-2}\,\mathrm{Hz}$ and achieves residual noise comparable to the glitch-subtracted L2 reference between $1\times10^{-3}$ and $6\times10^{-2}\,\mathrm{Hz}$. These results support subtracting glitches before transfer-function estimation and correlated cross-talk subtraction for this dataset.

\section{conclusions}
\label{sec:conclu}

In this work, we implemented and validated a transfer-function-based noise subtraction method using the coherence between the science channel and auxiliary channels, and applied it to cross-talk noise subtraction in the LPF science channel. 

Using simulated data, we validated the performance of the proposed transfer-function-based noise subtraction method and established a practical framework relating the auxiliary-channel noise level to the achievable subtraction performance. The results show that lower auxiliary-channel noise levels enable more accurate estimation of the common disturbance and hence more effective noise subtraction, thereby providing quantitative guidance for auxiliary-channel noise requirement design in future space-based GW detectors.

We then applied the proposed method to the public LPF telemetry data using 9 position and attitude auxiliary channels from the OMS and the GRS. 
For a 13-day dataset taken under stable conditions, it was found that transfer functions estimated from the full 13-day interval and separately for each day yield nearly identical subtraction performance over the frequency range under consideration, supporting the use of full-interval estimates for this dataset. In this case, the proposed method achieves subtraction performance comparable to that of the standard LPF fitting-based method in the frequency band $[1\times10^{-2},6\times10^{-2}]\,\mathrm{Hz}$~\cite{Wanner_2017}. At higher frequencies, it yields lower residual noise than the standard fitting-based result. The effect of the glitch subtraction was also studied, demonstrating that subtracting glitches before the transfer-function estimation can reduce the residual noise over part of the considered frequency range.

The results demonstrate the potential of transfer-function-based noise subtraction for data analysis in mHz space based GW detectors. The method can be extended to other LPF auxiliary channels and, in principle, to LISA. Its effectiveness depends on the stability of the coupling over the analysis interval and on the ability of the auxiliary channels to trace the target noise. Adding unrelated channels or channels with substantial intrinsic noise may provide no additional benefit and could diminish the subtraction power. 

\begin{acknowledgments}
This research is supported by the National Natural Science Foundation of China (NSFC) under Grant No.~12547104. 
We acknowledge the use of public data from the LISA Pathfinder mission, provided by the European Space Agency (ESA) and the LISA Consortium~\cite{LPF_Legacy_Archive}.

\end{acknowledgments}

\section*{Data Availability}

The LISA Pathfinder data used in this work are publicly available from the LISA Pathfinder Science Archive hosted by the European Space Agency (ESA)~\cite{LPF_Legacy_Archive}.


\bibliography{main}

\end{document}